**Precise phase control of large-scale inorganic perovskites via vapor-phase anion-exchange strategy**

*Guobiao Cen[1], Yufan Xia[1], Chuanxi Zhao[1]\*, Yong Fu[1], Yipeng An[2], Ye Yuan[1], Tingting Shi[1]\*, Wenjie Mai[1]\**

[1]Siyuan Laboratory, Guangdong Provincial Engineering Technology Research Center of Vacuum Coating Technologies and New Energy Materials, Department of Physics, Jinan University, Guangzhou, Guangdong 510632, People's Republic of China.
[2]School of Physics & International United Henan Key Laboratory of Boron Chemistry and Advanced Energy Materials, Henan Normal University, Xinxiang, Henan 453007, China.

E-mail: tcxzhao@email.jnu.edu.cn, ttshi@email.jnu.edu.cn, wenjiemai@email.jnu.edu.cn

Anion exchange offers great flexibility and high precision in phase control, compositional engineering and optoelectronic property tuning. Different from previous successful anion exchange process in liquid solution, herein, we develop a vapor-phase anion-exchange strategy to realize the precise phase and bandgap control of large-scale inorganic perovskites by using gas injection cycle, produing some perovskites such as $CsPbCl_3$ which has never been reported in thin film morphology. *Ab-initio* calculations also provide the insightful mechanism to understand the impact of anion exchange on tuning the electronic properties and optimizing the structural stability. Furthermore, because of precise control of specific atomic concentrations, intriguing tunable photoluminsecence is observed and photodetectors with tunable photoresponse edge from green to ultraviolet light can be realized accurately with an ultrahigh spectral resolution of 1 nm. Therefore, we offer a new, universal vapor-phase anion exchange method for inorganic perovskite with fine-tunable optoelectronic properties.

## 1. Introduction

Integration of photodetectors with gradient light-response offers a very promising prospect in the field of optoelectronic applications, such as spectrum imaging[1], telecommunications[2], mulitcolor detection[3], and artificial vision[4]. Conventional technology for fabricating color-selective sensors is usually based on complex color filter arrays[5] or assembled individual photodetectors[6]. However, considering the device complexity and high cost during the integration, some efficient stragtegies using composition engineering has been proposed. For example, Yang et al. reported compositionally graded semicondutor $CdS_xSe_{1-x}$ nanowire based single-nanowire spectrometers by thermal evaporation method, with spectral response along the length[1b]. Dou et al. successfully reported multicolor $CsPbX_3$ heterojunction nanowires via solution-phase anion exchange and electron-beam lithography[7]. Composition engineering is often used in the fabrication of photodetector to tune the optoelectronic properties[8]. Until now, the large-scale growth of patterned semiconductor films with spatial compositional gradients is still challenging by solution method, but it needs to be solved for fabricating high-integrity optoelectronic devices.

Inorganic halide perovskites ($CsPbX_3$, X= Cl, Br, I), as wide-band gap perovskites, are becoming the research spotlight due to their application in blue LEDs, photodetectors and multi-junction solar cells[9]. It is fast and facile to transport anion and exchange halide ions in inorganic $CsPbX_3$ because of its low defect formation energy, existentence of vacancies, as well as the labile nature of halide anions[10]. Therefore fast anion-exchange reaction in liquid phase highly facilitates the composition engineering of $CsPbX_3$ perovskites[11]. Additionally, anion exchange paves the way for fine-tunning the photoresponse range of perovskites, which can provide a powerful tool to realize color-selective sensors. Typical solution methods are post-synthesis procedure, which mixes perovskites with reactive anion precursors[12], [13]. However, currently the exchange reaction involving inorgainic perovskites has been realized successfully only in liquid solution[11a, 14]. The solution procedure often introduces capping ligands (such as oleic acid (OA), oleylamine (OLA)), which lead to high dynamic instability, form insulating layer on the surface and block the carrier transport[15]. Moreover, solution method is usually employed to synthesize nanocrystals[16] and quantum dots[17] of inorganic perovskites in small scale. So far, large-scale fabrication of some wide-band gap inorganic perovskite films (such as $CsPbCl_3$) has not been reported. The most crucial challenge is ascribed to the ultralow solubility of CsCl in inert solvents (such as dimethylsulfoxide (DMSO), N,N-dimethylformamide (DMF), Figure S1). The evaporation of low-solubility solvent leaves behind large amounts of grain boundaries and causes plentiful defects and phase segregation, which inevitablely hinder the charge transport. Previous studies indicate chlorine additives or precursors have an impact on morphology and surface defects but with little or no incorporation into perovskites[11b, 18]. Alloying high molar amounts of Cl into perovskite lattice becomes extremelly challenging, aiming for achieving high-efficiency wide-band gap perovskite tandems[19]. Furthermore, the preliminary understandings about anion exchange of perovskites are reported[20], but few investigation focused on mechanism into phase transformation based on first-principle calculations[21], which is of fundamental importance to unveil the structure-property relationship.

Herein, vapor-phase anion exchange strategy has been employed for the first time to precisely control the phase engineering of inorganic perovskites. Our technique provides unique high-reactive chemcial vapor, which skillfully avoid the drawbacks of solutibility and defect. Meanwhile, vapor phase precise control method such as atomic layer deposition (ALD) has been proved to be an ideal interface engineering technique to passivate defect states, enhance stability, as well as improve charge extraction[22]. Significantly, the proposed vapor-phase anion-exchange reaction in a chamber is proved to be a general strategy for many inorganic pervosksites, where the anion-exchange rate and reaction extent can be delicately controlled. The absorption spectral resolution can be precise controlled as low as 0.8 nm by gas injection cycles and reaction temperature. In order to unveil the insightful mechanism further, Density Functional Theory (DFT) calculations are also used to investigated the results which are consistent with the experimental characterizations of anion-exchanged perovskite. By precisely controlling the localized anion exchange reaction, spatially resolved multicolor CsPbX$_3$ perovskite (X=Cl, Br, or alloy of two halides) heterojunctions with well-defined patterns were fabricated. Finally, self-powered, tunable photoresponse, highly stable PDs were fabricated, showing high sensitivity from ultraviolet to visible region (3.28×10$^{12}$ Jones @ 405 nm), ultralow dark current (10$^{-11}$ A), ultralow UV detection limit (2.57 nWcm$^{-2}$), as well as high spectral resolution.

## 2. Results

### 2.1. Anion-exchange Reaction and Optical Characterization

We adopt novel vapor-phase method for large-scale anion exchange of uniform and compact CsPbCl$_x$Br$_{3-x}$ thin films, which avoids the drawbacks of solution methods. Typcial chemical vapor anion-exchange process is schematically illustrated in Figure 1a. Firstly, a uniform CsPbBr$_3$ thin film with thickness of ~80 nm was spin-coated on FTO substrates, and they were subsequently transferred into a chamber with presetted paremeters (105 °C, 9 Pa). Secondly, the Cl ion would be disassociated when the precursor of TiCl$_4$ purged into the chamber, along with fast anion exchange reaction of Br. Thirdly, the composition and bandgap of CsXBr$_3$ thin films can be continuously tuned by selected area exchanges using facile shadow masks. The extent of the anion-exchange reaction can be controlled by gas injection cycle of precursor. To show the chemical composition distribution, we performed elemental analysis of anion-exchanged perovskite films by transmission electron microscopy-energy dispersive X-ray spectroscopy (TEM-EDS). Figure 1b-d shows the morphology and EDS mapping of pervoskite thin films treated with 0, 20 and 50 gas injection cycles, respectively. In Figure 1b, uniform distribution of Cs, Pb and Br elements on the surface of as-fabricated CsPbBr$_3$ sample. After underwent 20 cycles of chemical vapor anion exchange,

Br would be partially subtituted by Cl ions, which is corresponding to the mixed phase $CsPbCl_xBr_{3-x}$ ($CsPbX_3$ alloy), as shown in Figure 1c. Further extending gas injection cycle to 50, the mixed phase would be almost conversed to $CsPbCl_3$ phase with minor Br left. Similar chemical composition exchanges were also confirmed by corresponding SEM-EDS mappings (Figure S2). Furthermore, the structure evolution of $CsPbBr_3$ films under different gas injection cycles was characterized by ex-situ XRD patterns, as shown in Figure 1e. For the as-prepared $CsPbBr_3$ films, the diffraction peaks locate at 15.18º, 21.45º, 30.66º, corresponding well to (100), (110) and (200) planes of monoclinic phase of $CsPbBr_3$ (JCPDS#18-0364, a=b=5.827 Å, c=5.891 Å, α=β=90º, γ=89.65º), respectively. After 10 cycles of anion exchange, all the diffraction peaks redshift toward larger angle direction. Further increasing gas injection cycle to 20, more intense redshift can be found, while the FTO diffraction peaks remain unchanged. When the gas injection cycles increase to 50, three diffraction peaks have moved to locations at 15.8º, 22.46º, and 32.0º which can be well indexed to (100), (101), and (200) facets of tetragonal phase $CsPbCl_3$ (JCPDS#18-0366, a=b=5.584 Å, c=5.623 Å, α=β=γ=90º). Here, we concentrate on analyzing the highest diffraction peak shift at ~21.47º, which is assigned as (110) facet of $CsPbBr_3$. From 0 to 50 gas injection cycles, the redshift of diffraction peak at 22.43º indicating the full substitution of $Br^-$ with $Cl^-$. Figure 1f shows the steady-state photoluminescence (PL) spectra of $CsPbCl_xBr_{3-x}$ thin films under 365 nm excitation. Inset displays typical photographs of $CsPbCl_xBr_{3-x}$ thin films when 365 nm UV-LED beams illuminate the samples. The emission peak continuously blue shifts to shorter wavelength with the increasing of gas injection cycle, confirming the controllable anion-exchange reaction and band gap change. Significantly, when gas injection cycle increases to 50, the emission peak of perovskite films shifts from 525 nm to 412 nm, and the latter peak is corresponding to the excitonic signature of intrinsic $CsPbCl_3$ perovskites[23]. Since Cl ion has smaller atomic radius, the Br ion in halide perovskites would be easily replaced via chemical vapor anion-exchange reaction. The gradual substitution of Br$^-$ with Cl$^-$ in perovskites lattice is accounting for the shifts of diffraction peaks and PL peaks.

Next, to identify the chemical composition of anion-exchanged halide perovskites, high-resolution X-ray photoelectronic spectroscopy (XPS) spectra are measured. Figure 2(a-b) shows the high-resolution Br 3d and Cl 2p core level XPS spectra, respectively. After careful de-convolution, the Br 3d and Cl 2p core level XPS peak can be fitted into two peaks at 68.38/69.38 eV and 197.98/199.58 eV, respectively. It is interesting to note that, the doublet intensity of Br 3d XPS peak decreases with the increase of gas injection cycle. On the contrary, the converse trend is found for doublet Cl 2p core level XPS peak. In the meantime, the Cs 3d and Pb 4f XPS spectra keep unchanged (Figure S3a,b). More specifically, by quantifying the intensity of the Br 3d and Cl 2p XPS peaks, their contents varied with the gas injection cycle are extracted. With the increase of gas injection cycle, the ratio of Br/Cl for 10 cycles, 20 cycles, 50 cycles, 100 cycles are determined to be 1:0.83, 1:1.5, 1:9, 1:61, respectively. This comparison strongly supports that Br can be continuously exchanged by Cl during the gas injection cycling process. It's worth noting that the original CsPbBr$_3$ can in situ transforms into CsPbCl$_3$ after 50 gas injection cycle of anion exchange. Figure 2c shows the tendency of Br/Cl atom ratio in the mixed CsPbCl$_x$Br$_{3-x}$ films, which clearly confirms the substitution can be precisely tuned by gas injection cycle. The whole anion-exchange evolution can be explained by the process of Cl$^-$ replacing Br$^-$ under low temperature. To figure out the element distribution in anion-exchanged perovskites, XPS depth profiling was used to analyze the chemical composition changes as a function of depth (50 cycles sample). As shown in Figure S3c, besides the presence of Cs, Pb, Cl and Br content, there also exists Ti, O, C, Sn element (Sn and O are from the FTO substrate). It indicates that the content of Ti increases with the increase of gas injection cycle (Figure S3d). To determine whether the Ti content would contribute to the anion-exchange, the Ti content as the function of etching depth was explored. Figure 3d shows the comparison of Ti$_{2p}$ XPS spectra on the surface and inner of sample, the largely attenuated XPS intensity in depth of 3.4 nm indicating Ti only exists on the surface of perovskites films, which in turn serves as a protection layer and contributes to the high stability of as-fabricated perovskite devices.

There is a pressing need for fabrication of uniform and high-quality bandgap tunable perovskite thin films, which is the base for high-performance thin film optoelectronic devices. The proposed vapor phase anion-exchange method can well take the above challenges for device fabrication. Figure S4 show typical SEM images of $CsPbBr_3$, mixed $CsPbCl_xBr_{3-x}$ and $CsPbCl_3$ perovskite thin films. After 50 cycles of anion exchange, the mixed $CsPbCl_xBr_{3-x}$ films become uniform and compact with less pinholes. However, more pinholes with smaller size would introduce after 100 gas injection cycles (Figure S4e). Based on the above structural, optical and morphology analysis, the optimal gas injection cycle for fabrication of high-quality $CsPbCl_3$ perovskite films is suggested to be 50. Therefore, gas injection cycle is set to be 50 for the following fabrication of UV photodetectors, which we will discuss later.

**2.2 DFT Calculation and Band Structure Analysis for Anion-exchange Mechanism**

To unravel the anion exchange mechanism accounting for the above structural, optical and morphology change, as well as its influence on the stability, we further used Vienna ab initio simulation package (VASP) based on density functional theory (DFT) to investigate the electronic properties systemetically.[24] In Figure 3a, Figure 3b and Figure 3c, the projective Heyd-Scuseria-Ernzerhof (HSE03) band structures of $CsPbBr_3$, $CsPbBr_{1.5}Cl_{1.5}$ and $CsPbCl_3$ are calculated respectively.[25] It is obvious to see that the band gap increases from 2.28 eV to 2.79 eV when the structure changes from $CsPbBr_3$ to $CsPbCl_3$ at α-phase (Pm-3m). The total density of states (TDOS) and band gaps of $CsPbBr_3$, mixed $CsPbCl_xBr_{3-x}$ and $CsPbCl_3$ perovskite thin films can be found in Figure 3a-c, respectively. Figure 3d shows the intriguing trend of bandgap change under different ratio of chlorine element. Herein, we applied three calculation methods for the high precision, including generalized gradient approximation (GGA),[26] hybrid functional Perdew-Burke-Ernzerh (PBE0)[27] with Spin Orbital Coupling (SOC)[28] and HSE03. Details of DOS and projecive band analysis can be found in Figure S5-11. All the calculated results are in line with the experimental observations, indicating the similar trend of electronic properties. Among all the calculations, HSE03 method shows the closest bandgap values with the experimental ones, which agrees well with the above observations. Our GGA calculations is also consistent with the previously reported bandgaps of $CsPbX_3$ (as shown in Table S1).[29] To better understand the electronic band structure after anion exchange, we use the schematic diagram to elucidate the bandgap increases with the bond lengths of Pb-X (inset of Figure 3d). When the Br atoms are gradually substituted by Cl atoms, the band gap of mixed halide perovskites will increase due to the longer bond length of Pb-Br (2.995 Å) than that of Pb-Cl (2.826 Å). As shown in the inset of Figure 3d, after Cl atoms substituting Br atoms, the gap value between conduction band minimum (CBM) and valence band maximum (VBM) will be enlarged, and the band gap begins to broaden. Hence, the possibility of <p|p> transition will depand on higher energy absorption due to wider band gap value in $CsPbCl_3$ perovskite. Additionally, to explore the stability of transformed $CsPbBr_3$ perovskites, the Goldschmidt tolerance factor (t) before and after Cl substitution is also calculated by the following equation[30]: $t=(r_A+r_X)/\sqrt{2}(r_B+r_X)$, where $r_A$ is the radius of the A cation, $r_B$ is the radius of the B cation, and $r_X$ is the radius of the anion. The t value gets closer to 1, the better stabillty of perovskites. As shown Figure 3e, the t value increases from 0.862 to 0.866, and 0.870 for $CsPbBr_3$, $CsPbCl_{1.5}Br_{1.5}$ and $CsPbCl_3$, respectively. According to the reliable empirical index of t, the structure of transformed $CsPbCl_xB_{X_{3-x}}$ perovskite films is reasonable to show higher stability than that of as-grown $CsPbBr_3$ perovskite films. Our result is also consistent well with recent report which found chlorine mixing would enhance the stability of hybrid perovskites[21].

**2.3 Localized Anion-exchange Reaction and Confocal PL Mapping**

Since the vapor-phase anion-exchange process is based on chemical diffusion, which is benificial for large-scale transformation of perovskite thin films. The anion diffusion is chemical vapor-exposure determining step that gives rise to the gradual transformation to heterogeneous Br/Cl mixed inorganic perovskites and ultimately to $CsPbCl_3$ phase with wide band gap, enabling the anion-change propagation to be visualized with varied stage of gas injection cycle. Significantly, it is of great challenge to realize microscale control of the anion exchange at particular position. As a proof-of-concept study, we further employ a precision ceramic shadow mask (pore diameter $\varphi=160$ μm) to realize microscale control of anion-exchange reaction in $CsPbBr_3$ perovskite films. Figure 4a shows the schematic illustration of the chemical-vapor anion-exchange process. The as-formed $CsPbBr_3$ pervoskite films as the subjacent layer is covered by hard ceramic mask with pressure. To demonstrate the continuous stages of anion-exchange with perovskite films, we further use shadow mask by high-temperature tape on top of the ceramic mask (each piece of tape covering three columns of micropores), as shown in Figure 4b. Four different gas injection cycles (5, 15, 25, 50) are chosen, as noted I, II, III, and V, respectively. The left image shows photograph of ceramic mask, and the right image shows optical image of ceramic mask under microscopy. Figure 4c shows the fluorescence microscopy images, which can record the corresponding Cl-exchanged areas. For gas injection cycle of 5, the shadow feature of ceramic mask can be clearly identified, just like traditional lithography technology. The non-exposure area (label 1) in the center exhibits green color which is corresponding to the intrinsic color of $CsPbBr_3$. As label 2 in Figure 4c, the anion-exchanged area exhibits light-blue color, indicating partial Br substitution by Cl atoms. With the increase of gas injection cycle to 15, the profile and diameter of Cl-exchanged micropores become clearer and larger, respectively. When the gas injection cycle further increase to 50, the color of anion-exchanged micorpores would change to dark blue. The PL spectra also indicate the sucessful anion-exchange process by ALD-chemical vapor. Figure 4d shows the diameter of micropores approximately linearly increases with the cycle number, indicating the precise control of the anion exchange process. Additionally, PL patterns of "JNU" letters illuminated by 365 nm UV LED can be clearly observed, as shown in Figure 4e. The area dark on the top is due to the tapped FTO substrate. In order to realize continuous PL images in one sample, perovskite stripes with varied gas injection cycles have been demonstrated by using high-temperature tape mask (stripe width ~1 mm). As shown in Figure 4f, it can be seen that the strip color can be continuously tuned from green to light green, dark green, then to dark blue, purple, corresponding to 0, 5, 10, 20, 30 gas injection cycles, respectively.

## 2.4 Wide-range and Fine-tuning Optical Absorption and General Strategy for Other Inorganic Pervoskites

Figure 5a shows the optical absorption spectra of mixed-halide perovskite films varied with gas injection cycle. The intrinsic absorption peaks show narrow and pronounced excitonic band feature, and the absorption edges blue shift from 532 nm to 416 nm with the increasing of gas injection cycle from 0 to 50 cycles. The band gaps of $CsPbCl_xBr_{3-x}$ films are calculated to be 2.33, 2.51, 2.55, 2.71, 2.82 eV and 2.98 eV, respectively (Figure S12). Benifitting from the flexibilities in controlling of experiment parameters, such as reaction temperature, gas flow rate and gas injection cycle number, the absorption peak of photoresponse layer ($CsPbCl_xBr_{3-x}$) can be precisely controlled. In order to deterimine the spectral resolution limit, we slow down the reaction speed by lowering the reaction temperature to room temperature. As shown in Figure 5b, the peak locates at 516.8 nm corresponding to the intrinsic absorption band of $CsPbBr_3$. Under given pulse time (8 ms), the absorption peak of $CsPbCl_xBr_{3-x}$ pervoskite shifts from 515.2 nm to 513.2 nm by ultilizing different gas injection cycle (2~30 cycles). The corresponding offsets are 1.6 nm, 1.2 nm and 0.8 nm, respectively. Increasing the gas injection cycle further, the absorption peak become relative stable with minor change for the low reaction temperature. After rasing the reaction temperature to 60 ºC, the adsorption peak further blue shifts to 512.2 nm. Considering the wavelength accuracy (±0.3 nm) of spectrometer, the spectral resolution limit by vapor strategy can be reasonablely controlled below 1 nm. To verify the effectiveness of chemical-vapor anion-exchange reaction on other inorganic pervoskites, typical Pb-based pervorskite ($CsPbIBr_2$), Bi-based pervorskite ($Cs_3Bi_2Br_9$), and double pervoskite ($Cs_2AgBiBr_6$) were choosen. As shown in Figure S13(a-c), the absorption spectra of three pervoskite thin films after anion exchange are compared. All the absorption edges clearly blueshift to shorter wavelength after certain ALD-assisted conversion. It can be seen that $CsPbIBr_2$ pervoskite shows the largest shift (~18 nm) among them, while the $Cs_2AgBiBr_6$ exhibits the smallest shift (<3 nm), even after 50 gas injection cycles. The offset $\Delta\lambda$ also indicates the stablity of inorganic pervoskites, which can experience anti-anion exchange reaction during the chemical process. Based on the above analysis, our anion-exchange method is still a general strategy for effective, precise control of the anion exchange in inorganic pervoskites. The incorporation of Cl into bulk lattice would reduce the defect density, increase carrier mobility, as well as increase the stability [19, 21].

**2.5 Near Ultraviolet Photodetection and Tunable Photoresponse**

We have proved the high-effectivity of vapor phase for bandgap engineering of inorganic perovskites. By assistance of deposition technique, ultrathin $Al_2O_3$ and $TiO_2$ modification layers are incorporated into the interfaces between FTO and perovskites to fabricate better-performance photodetectors. Figure 6a shows typical photoresponse curves of $CsPbCl_3$ photodetectors measured under 405 nm laser illuminations with different power density. The heterojunction photodetectors exhibit remarkable switching characteristic between high and low conduction mode under modulated light (405 nm, 0.25 Hz). As show in Figure 6b, the photocurrent shows almost linear dependence on the light intensity in the range from 2.57

nWcm$^{-2}$ to 5 mWcm$^{-2}$. The power law I∝P$^{\alpha}$ was used to fit the I-P curve, and the ideal index can be calculated to be 0.89. The fitting factor is close to the ideal value of 1, implying the high-quality CsPbCl$_3$ thin films have low trap states. The linear dynamic range (LDR) can be determined to be 125 dB by the equation: LDR=20log(P$_{max}$/P$_{min}$), where P$_{max}$ and P$_{min}$ are the maximum and minimum optical power levels measured by the PD. Benefiting from the double-side modification layers, the dark current of FTO/Al$_2$O$_3$/CsPbCl$_3$/TiO$_2$ (ACT) PDs can be suppressed to the lowest level of 10$^{-11}$ A, along with the detectable limit of 2.57 nWcm$^{-2}$, as shown in Figure S14a. Besides, the UV photodetectors also exhibit fast response speed, with rise time and decay time of 7 ms and 165 ms, respectively (Figure S14b). The specific detectivity is an important parameters for evaluating the weak light detecting capability of ACT PDs, which can be calculated by D*=R/(2qJ$_d$)$^{1/2}$, where R is responsivity, and J$_d$ is dark current density. As plotted in Figure 6c, R and D* decrease almost linearly with the increase of light intensity in the range from 2.57 nWcm$^{-2}$ to 5 mWcm$^{-2}$. The highest R of 0.1 AW$^{-1}$ can be obtained. The specific detectivity D* is determined to be as high as 3.28×10$^{12}$ Jones (@405 nm). Furthermore, the spectral response curves of perovskite PDs with different gas injection cycles were also measured. As shown in Figure 6d, the CsPbBr$_3$ PDs show a broader response range with high sensitivity located at 520 nm, which corresponding to the adsorption edge of CsPbBr$_3$ (Figure 5a). After 10 cycles of ALD-assisted anion exchange, partial Br atoms would be replaced by Cl atoms and forming mixed perovskites, the photoresponse edge would blueshift to 490 nm for the broadening band gap. When the gas injection cycle increases to 20 and 50, the photoresponse edge would further blueshifts to 460 and 429 nm, respectively. Based on aforementioned adsorption measurement, the mixed pervoskites would fully converse to CsPbCl$_3$ after 50 gas injection cycles. Therefore, the response spectrum range corresponds to the absorption band gap of CsPbCl$_3$. It's worth mentioning that, the photoresponse range can be easily tuned from visible to ultraviolet by precise control of the gas injection cycle. The interval of the cutoff response edge is around 30 nm which is associated with above adsorption spectra, revealing a high spectra resolution of ~30 nm can be achieved using the devices. The wafer area could be easily scaled up by the anion-exchange method, together with interfacial engineering, strengthen the competitiveness of all inorganic perovskite-based UV optoelectronic applications.

## 3. Conclusion

In this work, through novel in situ conversion method, we have successfully established a universal low-temperature chemical-vapor anion-exchange strategy for exchange halogen anions in inorganic perovskites. During the vapor-phase reaction process, Br atoms can be gradually substituted by Cl atoms, which have been confirmed by optical absorption, XRD and XPS characterization. DFT calculations is employed to unveil the experimental results and show the consistent trends on tuning the optoelectronic properties by changing the halide anions. Significantly, high-quality and large-scale uniform $CsPbCl_3$ perovskite thin films are obtained for the first time. Tunable photoresponse from green to near ultraviolet light ($\Delta\lambda$~100 nm) can be realized by our strategy, along with a remarkable spectral resolution of ~1 nm. Our general method can be extended to other inorganic pervoskites including Pb-based pervorskite ($CsPbIBr_2$), Bi-based pervorskite ($Cs_3Bi_2Br_9$), and double pervoskite ($Cs_2AgBiBr_6$). The wafer-area anion-exchange method, together with interfacial engineering, would strengthen the competitiveness of perovskite photodetectors with tunable response range, which has great potential application in the fields of machine vision and artificial vision.

## 4. Methods

Materials synthesis: The FTO glass substrates were sequentially washed with isopropanol, acetone, ionized water and isopropanol for 15 min, respectively. The sheet resistance of FTO is 15 $\Omega\square^{-1}$. After that, the cleaned substrates were dried by nitrogen flow, and following was treated by UV-ozone cleaner for 30 min. Then, the as-prepared substrates were put into atomic layer deposition system (ALD, Beneq TFS200) for following ALD-modification layer deposition. Aluminium trimethide (TMA) and H$_2$O were precursors, and high purity nitrogen was used as the carrier gas. The thickness of ALD-layer can be precisely controlled by adjusting the cycling number (growth rate=0.09 nm/cycle). For the growth of ALD-Al$_2$O$_3$ layer, the cycling number is set to be 16 under 85 $^o$C for 5 min, corresponding to an optimal thickness of 1.5 nm [22]. After that, the CsPbBr$_3$ thin films were spin-coated on the ALD-Al$_2$O$_3$ modified FTO substrate at room temperature. First, 0.4 mol CsBr and 0.4 mol PbBr$_2$ were mixed and dissolved in 1 mL of dimethyl sulfoxide (DMSO) solution, which was stirred on a hot plate (70 $^o$C).

Vapor-phase halogen exchange: Afterward, the precursor solution was spin coated onto the as-prepared substrates (FTO and ALD-Al$_2$O$_3$ coated FTO), the spin-coating parameters are set to be 500 rpm for 6 s and 4000 rpm for 30 s. The vapor-phase halogen reaction was performed in chamber (105 $^o$C, 9 Pa). Titanium tetrachloride (TiCl$_4$) as precursors were purged on perovskites surface, and high purity nitrogen was used as carrier gas. The content of TiCl$_4$ gas per cycle is 200 ppm, and the N$_2$ purge time is 2s.

Devices fabrication: After the synthesis of CsPbBr$_3$ thin films, ultrathin layer of TiO$_2$ were further deposited on the CsPbBr$_3$ perovskite thin films by ALD technique. Titanium tetrachloride (TiCl$_4$) and H$_2$O were precursors, and high purity nitrogen was used as carrier gas. The growth rate was 0.06 nm per cycle at 105 $^o$C for 17 min. Thickness of TiO$_2$ layers were controlled by adjusting the cycle numbers. After the deposition of dense and compacted TiO$_2$ layers, 100 nm Au electrode was deposited on the front of TiO$_2$ layer by thermal evaporation method (4×10$^{-4}$Pa).

Materials characterization: The structure for CsPbBr$_3$ perovskite thin films was characterized by X-ray diffraction (XRD) technique (Rigaku, Miniflex600). The energy band alignments are measured by ultraviolet photoelectron spectra (UPS, He I radiation, hν=21.22 eV). All the samples were biased with -5 V to obtain the low energy secondary electron (SE) cutoffs. The optical absorption spectra of CsPbBr$_3$ perovskite films were characterized by UV-Vis spectrophotometer (SHIMADZU, UV-2600). Photoluminescence (PL) spectra of $CsPbBr_3$ perovskite thin films were obtained from photoluminescence spectroscopy (SHIMADZU, RF-5301PC). The SEM images of $CsPbBr_3$ perovskite film were obtained by using scanning electron microscope (SEM, ZEISS ULTRA 55). PL mapping was visualized using a Zeiss confocal fluorescence microscope (ZEISS, LSM700) and quantified with software.

DFT calculation: Density Functional Theory (DFT) Calculations. The calculation based on density functional theory (DFT) was carried out using Vienna ab initio simulation package (VASP).[24] Herein, we used the projector augmented waves (PAW)[31] as the pseudopotentials. The band structures and Density of States (DOS) were calculated by employing the Perdew–Burke-Ernzerh (PBE)[26] of the generalized gradient approximation (GGA) as the exchange-correlation functional. In addition, four high-symmetric k-points: Γ (0, 0, 0), X (0, 1/2, 0), M (1/2, 1/2, 0), and R (1/2, 1/2, 1/2) were included when we calculate the band structures. We chose 6×6×6 k-point mesh for the Brillouin zone.[32] The plane-wave cutoff energy of wave function was set as 400 eV. A 2×2×2 supercell including 40 atoms is used for the $CsPbBr_{1.5}Cl_{1.5}$ calculations. To get the accurate bandgap, we used the Heyd-Scuseria-Ernzerhof (HSE) hybrid functional[25] containing HSE03 and HSE06 for the further calculation. Meanwhile, we utilized the hybrid functional PBE0[27] and found the similar band structures with overestimated values. Noticeably, all our calculations show the same trend of band gaps, especially for HSE03 method, whose theoretical band gaps are in good agreement with the experimental values.

Photoresponse measurement: To evaluate the photoresponse performance of $CsPbBr_3$ perovskite PDs, the I-V curves and photoresponse curves were measured by Keithley source meter (2601A). The monochrome light is 405 nm laser sources, the intensity is calibrated by a standard Si power meter (LE-LPM-HS411). The time-dependent photocurrent curves were measured under illumination of 405 nm light. The output power could be adjusted and controlled by neutral optical attenuation plates. The transient photoresponse behavior of the PDsis measured using the photo-induced open-circuit voltage decay method under a pulsed laser (405 nm). Photoresponse speeds of the PDs were evaluated by combining a pulse laser and a digital oscilloscope (Tektronix MSO 3054, 500 MHz). The spectral response (300-1100 nm) curve of the PDs is measured using a QE-R external quantum efficiency instrument (Enlitech, Si detector S10-14010), and the photocurrent is recorded by a Keithley 2601A source meter.

**Supporting Information**

Supporting Information is available from the Wiley Online Library or from the author.

**Acknowledgements**

Guobiao Cen and Yufan Xia contributed equally to this work. This work was supported by the National Natural Science Foundation of China (Grant Nos. 61604061, 51772135, 11804117 and 11774079), the Natural Science Foundation of Guangdong Province, China (Grant Nos. 2019A1515010482, 2020A1515011377), the Fundamental Research Funds for the Central Universities (Grant No. 21618405). We also thank Prof. Dong Ma and Dr. Qikun Cheng, Gai Wang for performing the conformal PL imaging measurement at Jinan University, Dr. Huawei Song, Chunhua Su for performing the TEM measurements at Sun Yat-sen University.

Received: ((will be filled in by the editorial staff))
Revised: ((will be filled in by the editorial staff))
Published online: ((will be filled in by the editorial staff))

**References**

[1] a) X. Tang, M. M. Ackerman, M. Chen, P. Guyot-Sionnest, *Nat. Photonics* **2019**, 13, 277; b) Z. Yang, T. Albrow-Owen, H. Cui, J. Alexander-Webber, F. Gu, X. Wang, T.-C. Wu, M. Zhuge, C. Williams, P. Wang, A. V. Zayats, W. Cai, L. Dai, S. Hofmann, M. Overend, L. Tong, Q. Yang, Z. Sun, T. Hasan, *Science* **2019**, 365, 1017.
[2] C. Sun, M. T. Wade, Y. Lee, J. S. Orcutt, L. Alloatti, M. S. Georgas, A. S. Waterman, J. M. Shainline, R. R. Avizienis, S. Lin, B. R. Moss, R. Kumar, F. Pavanello, A. H. Atabaki, H. M. Cook, A. J. Ou, J. C. Leu, Y.-H. Chen, K. Asanović, R. J. Ram, M. A. Popović, V. M. Stojanović, *Nature* **2015**, 528, 534.
[3] R. D. Jansen-van Vuuren, A. Armin, A. K. Pandey, P. L. Burn, P. Meredith, *Adv. Mater.* **2016**, 28, 4766.
[4] J. Xue, Z. Zhu, X. Xu, Y. Gu, S. Wang, L. Xu, Y. Zou, J. Song, H. Zeng, Q. Chen, *Nano Lett.* **2018**, 18, 7628.
[5] F. Yesilkoy, E. R. Arvelo, Y. Jahani, M. Liu, A. Tittl, V. Cevher, Y. Kivshar, H. Altug,

*Nat. Photonics* **2019**, 13, 390.

[6] F. P. García de Arquer, A. Armin, P. Meredith, E. H. Sargent, *Nat. Rev. Mater.* **2017**, 2.

[7] L. Dou, M. Lai, C. S. Kley, Y. Yang, C. G. Bischak, D. Zhang, S. W. Eaton, N. S. Ginsberg, P. Yang, *Proc. Natl. Acad. Sci. U. S. A.* **2017**, 114, 7216.

[8] a) Y. Fang, Q. Dong, Y. Shao, Y. Yuan, J. Huang, *Nature Photonics* **2015**, 9, 679; b) J. B. Rivest, P. K. Jain, *Chem. Soc. Rev.* **2013**, 42, 89.

[9] T. Leijtens, K. A. Bush, R. Prasanna, M. D. McGehee, *Nat. Energy* **2018**, 3, 828.

[10] D. Zhang, Y. Yang, Y. Bekenstein, Y. Yu, N. A. Gibson, A. B. Wong, S. W. Eaton, N. Kornienko, Q. Kong, M. Lai, A. P. Alivisatos, S. R. Leone, P. Yang, *J. Am. Chem. Soc.* **2016**, 138, 7236.

[11] a) G. Nedelcu, L. Protesescu, S. Yakunin, M. I. Bodnarchuk, M. J. Grotevent, M. V. Kovalenko, *Nano Lett.* **2015**, 15, 5635; b) S. Dastidar, D. A. Egger, L. Z. Tan, S. B. Cromer, A. D. Dillon, S. Liu, L. Kronik, A. M. Rappe, A. T. Fafarman, *Nano Lett.* **2016**, 16, 3563.

[12] S. E. Creutz, H. Liu, M. E. Kaiser, X. Li, D. R. Gamelin, *Chemi. Mater.* **2019**, 31, 4685.

[13] D. Parobek, Y. Dong, T. Qiao, D. Rossi, D. H. Son, *J. Am. Chem. Soc.* **2017**, 139, 4358.

[14] a) Y.-C. Chen, H.-L. Chou, J.-C. Lin, Y.-C. Lee, C.-W. Pao, J.-L. Chen, C.-C. Chang, R.-Y. Chi, T.-R. Kuo, C.-W. Lu, D.-Y. Wang, *J. Phys. Chem. C* **2019**, 123, 2353; b) K. Abdel-Latif, R. W. Epps, C. B. Kerr, C. M. Papa, F. N. Castellano, M. Abolhasani, *Adv. Funct. Mater.* **2019**, 29, 1900712.

[15] a) G. H. Ahmed, J. Yin, R. Bose, L. Sinatra, E. Alarousu, E. Yengel, N. M. AlYami, M. I. Saidaminov, Y. Zhang, M. N. Hedhili, O. M. Bakr, J.-L. Brédas, O. F. Mohammed, *Chem. Mater.* **2017**, 29, 4393; b) J. De Roo, M. Ibáñez, P. Geiregat, G. Nedelcu, W. Walravens, J. Maes, J. C. Martins, I. Van Driessche, M. V. Kovalenko, Z. Hens, *ACS Nano* **2016**, 10, 2071; c) A. Pan, B. He, X. Fan, Z. Liu, J. J. Urban, A. P. Alivisatos, L. He, Y. Liu, *ACS Nano* **2016**, 10, 7943.

[16] L. Protesescu, S. Yakunin, M. I. Bodnarchuk, F. Krieg, R. Caputo, C. H. Hendon, R. X. Yang, A. Walsh, M. V. Kovalenko, *Nano Lett.* **2015**, 15, 3692.

[17] W. Zheng, P. Huang, Z. Gong, D. Tu, J. Xu, Q. Zou, R. Li, W. You, J.-C. G. Bünzli, X. Chen, *Nat. Commun.* **2018**, 9, 3462.

[18] H. Tan, A. Jain, O. Voznyy, X. Lan, F. P. García de Arquer, J. Z. Fan, R. Quintero-Bermudez, M. Yuan, B. Zhang, Y. Zhao, F. Fan, P. Li, L. N. Quan, Y. Zhao, Z.-H. Lu, Z. Yang, S. Hoogland, E. H. Sargent, *Science* **2017**, 355, 722.

[19] J. Xu, C. C. Boyd, J. Y. Zhengshan, A. F. Palmstrom, D. J. Witter, B. W. Larson, R. M. France, J. Werner, S. P. Harvey, E. J. Wolf, *Science* **2020**, 367, 1097.

[20] a) D. Parobek, Y. Dong, T. Qiao, D. Rossi, D. H. Son, *J. Am. Chem. Soc.* **2017**, 139, 4358; b) Z.-J. Li, E. Hofman, A. H. Davis, M. M. Maye, W. Zheng, *Chem.Mater.* **2018**, 30, 3854.

[21] N. Rybin, D. Ghosh, J. Tisdale, S. Shrestha, M. Yoho, D. Vo, J. Even, C. Katan, W. Nie, A. J. Neukirch, S. Tretiak, *Chem. Mater.* **2020**, 32, 1854.

[22] G. Cen, Y. Liu, C. Zhao, G. Wang, Y. Fu, G. Yan, Y. Yuan, C. Su, Z. Zhao, W. Mai, *Small* **2019**, 15, 1902135.

[23] L. Protesescu, S. Yakunin, M. I. Bodnarchuk, F. Krieg, R. Caputo, C. H. Hendon, R. X. Yang, A. Walsh, M. V. Kovalenko, *Nano Lett.* **2015**, 15, 3692.

[24] G. Kresse, J. Furthmüller, *Phys. Rev. B* **1996**, 54, 11169.

[25] J. Heyd, G. E. Scuseria, M. Ernzerhof, *J. Chem. Phys.* **2003**, 118, 8207.

[26] J. P. Perdew, K. Burke, M. Ernzerhof, *Phys. Rev. Lett.* **1996**, 77, 3865.

[27] C. Adamo, V. Barone, *J. Chem. Phys.* **1999**, 110, 6158.

[28] a) J. Even, L. Pedesseau, J.-M. Jancu, C. Katan, *J. Phys. Chem. Lett.* **2013**, 4, 2999; b)

E. Narsimha Rao, G. Vaitheeswaran, A. H. Reshak, S. Auluck, *Phys. Chem. Chem. Phys.* **2017**, 19, 31255.

[29] M. Ahmad, G. Rehman, L. Ali, M. Shafiq, R. Iqbal, R. Ahmad, T. Khan, S. Jalali-Asadabadi, M. Maqbool, I. Ahmad, *J. Alloys Compd.* **2017**, 705, 828.
[30] Z. Li, M. Yang, J.-S. Park, S.-H. Wei, J. J. Berry, K. Zhu, *Chem. Mater.* **2016**, 28, 284.
[31] G. Kresse, D. Joubert, *Phys. Rev. B* **1999**, 59, 1758.
[32] J. D. Pack, H. J. Monkhorst, *Phys. Rev. B* **1977**, 16, 1748.

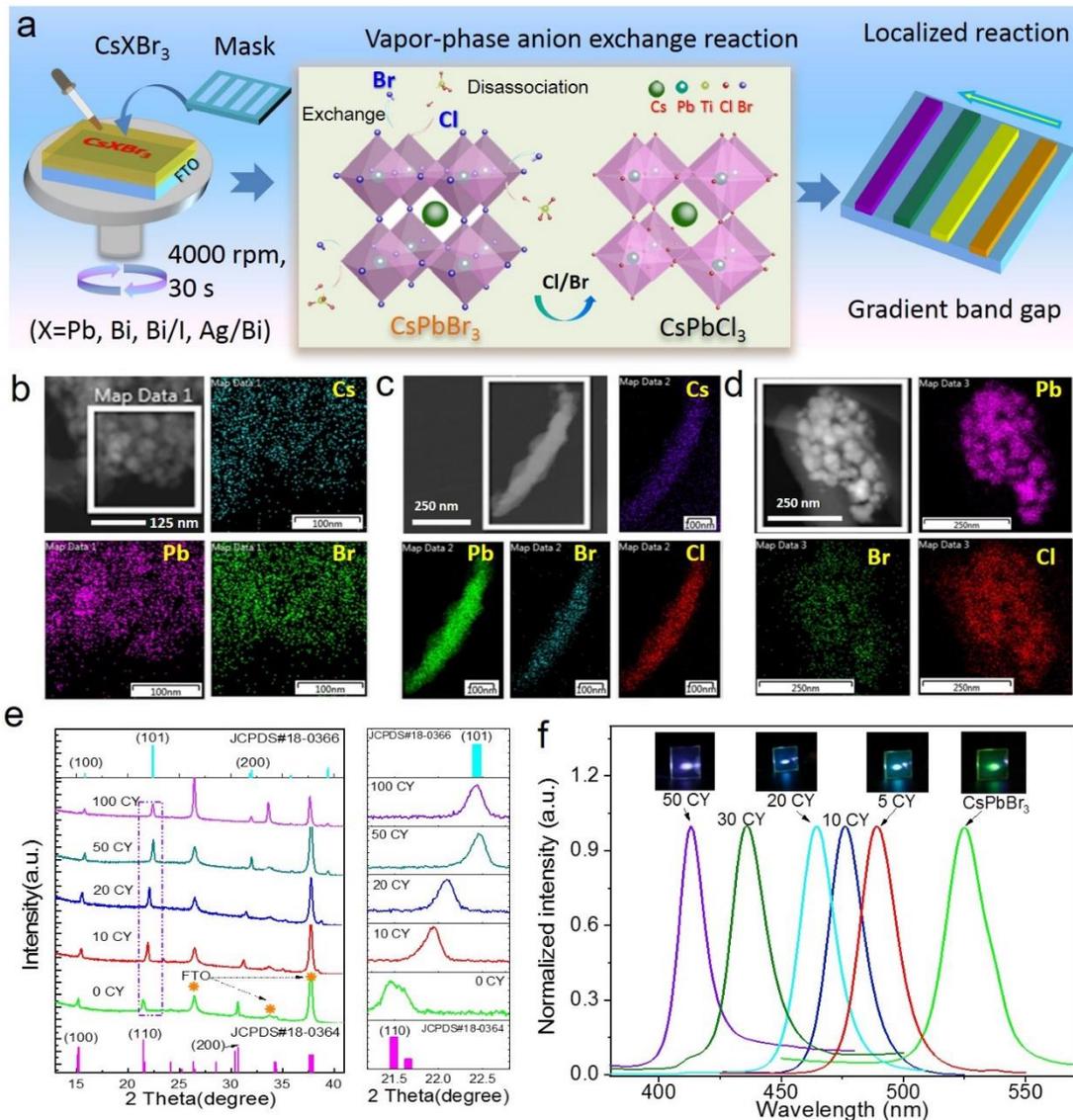

**Figure 1. Characterizations of chemical-vapor anion-exchange samples**. (a) Schematic of vapor-phase anion-exchange reaction, (b-d) TEM-EDS mapping of Cs, Pb, Br, Cl components of cesium lead halide perovskites with varied gas injection cycling number of 0, 20, and 50, respectively, (e) Comparison of XRD patterns of original CsPbBr$_3$ films and anion-exchanged CsPbCl$_x$Br$_{3-x}$ films, (f) Corresponding photoluminescence spectra. Insets of Figure 1g show the photos of the perovskite films under UV irradiation (λ=365 nm).

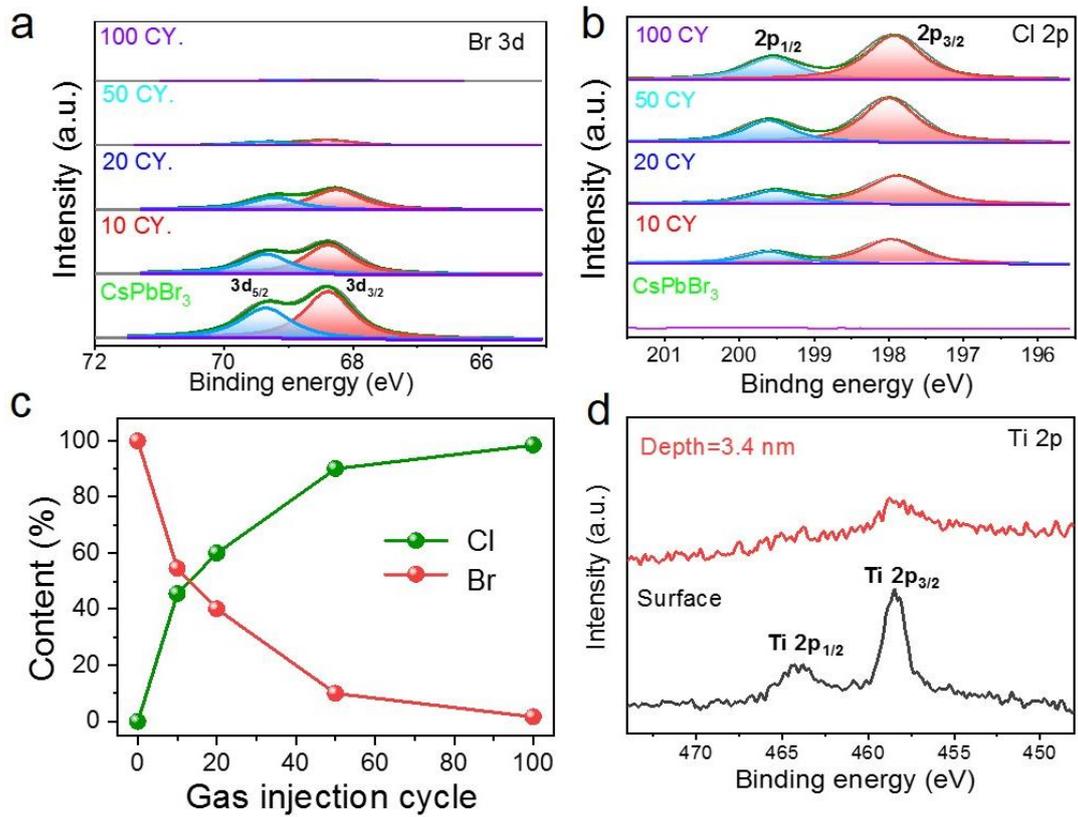

**Figure 2. XPS characterization of cesium lead halide perovskites with varied gas injection cycle**. (a) Br 3d core level XPS spectra, (b) Cl 2p core level XPS spectra, (c) Cl/Br content varied with the gas injection cycle, (d) XPS depth profiling of Ti 2p content on the surface and 3.4 nm below surface.

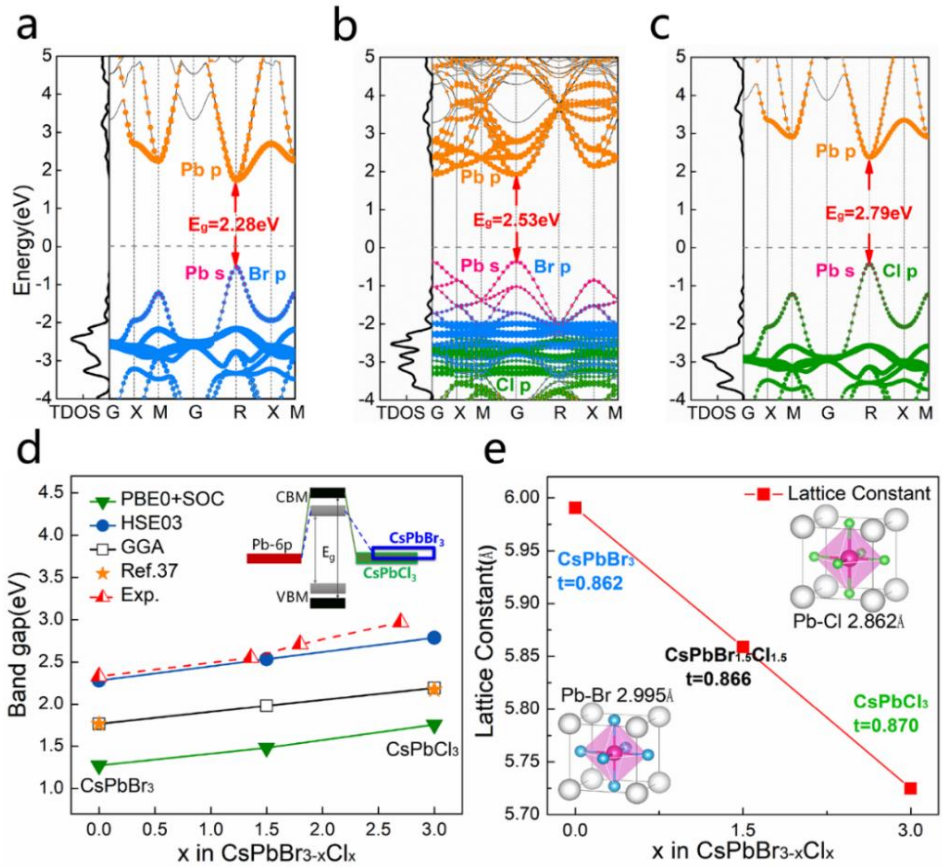

**Figure 3. DFT Calculation of Electronic Properties and Band-Trend Analysis.** DOS and projective band structure based on HSE03 method of (a) CsPbBr$_3$ perovskite, (b) CsPbCl$_{1.5}$Br$_{1.5}$ perovskite, (c) CsPbCl$_3$ perovskite, (d) Trend of calculated band gaps of CsPbCl$_x$Br$_{3-x}$ as a function of composition x of halide anions, inset gives the schematic diagram of band change due to anion exchange. (e) The calculated tolerance factor and lattice constant before and after Cl substitution.

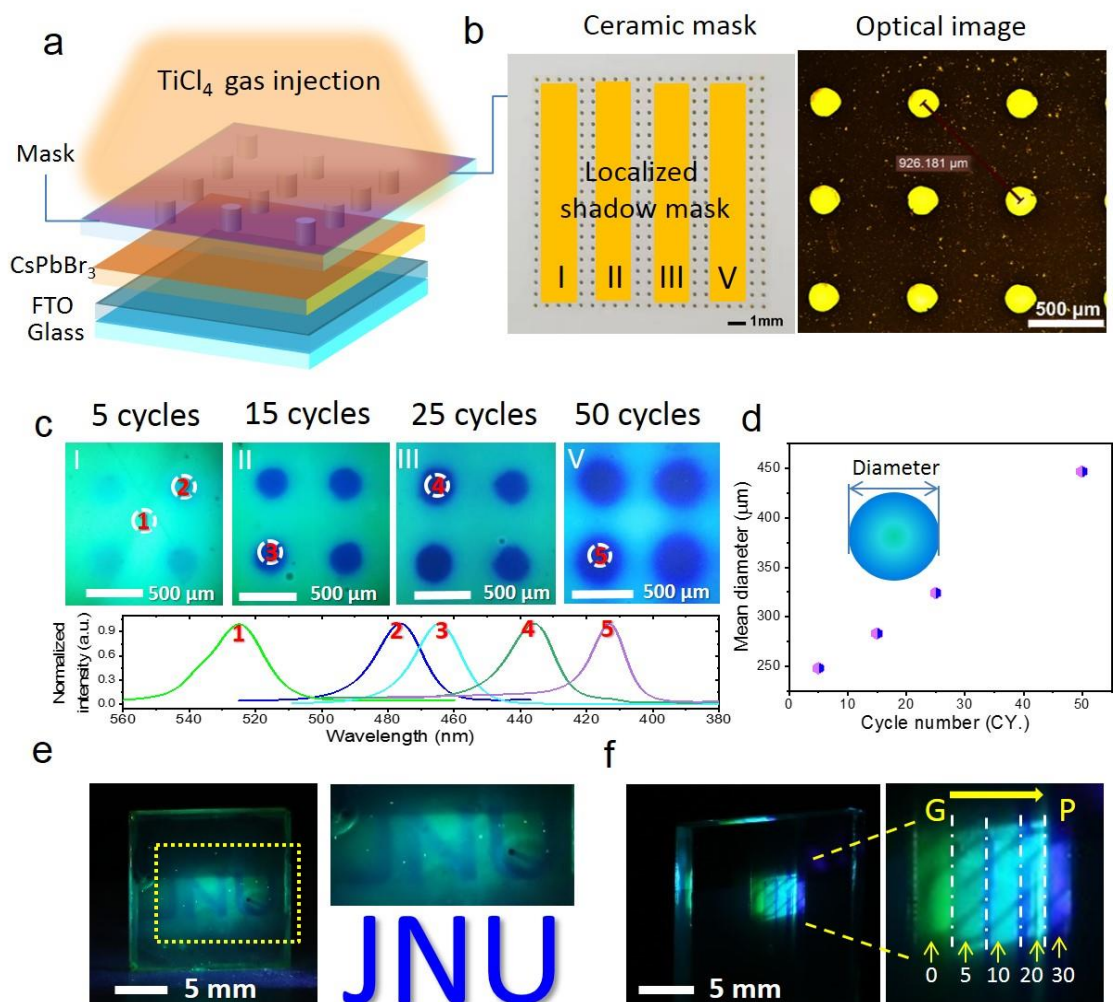

**Figure 4. Localized Anion-exchange Reaction and Confocal PL Mapping**. (a) A schematic illustration of the anion-exchange process. (b) ceramic mask and its optical image. (c) PL mapping of localized anion-exchanged areas with varied cycles. (d) Diameter of micropore varied with cyle number. (e) PL images of letters "JNU" and (f) strips under 365 nm excitation.

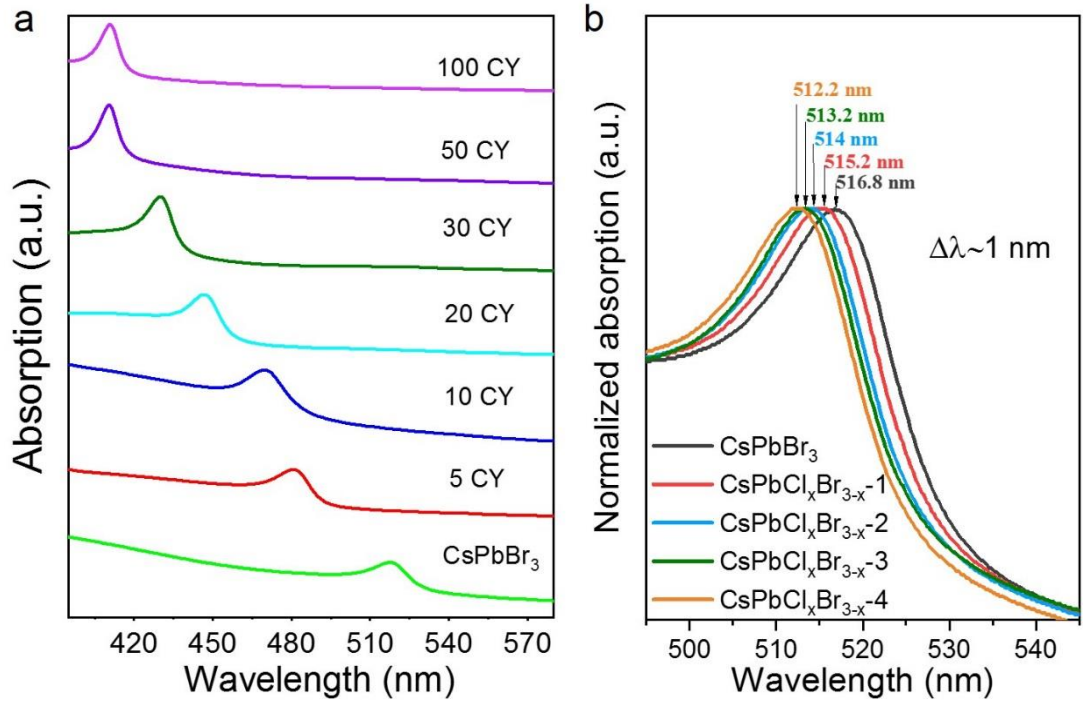

**Figure 5. Wide-range and Fine-tuning Optical Absorption of CsPbCl$_x$Br$_{3-x}$ pervoskites**. (a) Typical optical adsorption spectra by utilizing different cycles, (b) Fine-tuning adsorption spectra of CsPbCl$_x$Br$_{3-x}$ films under different cycles.

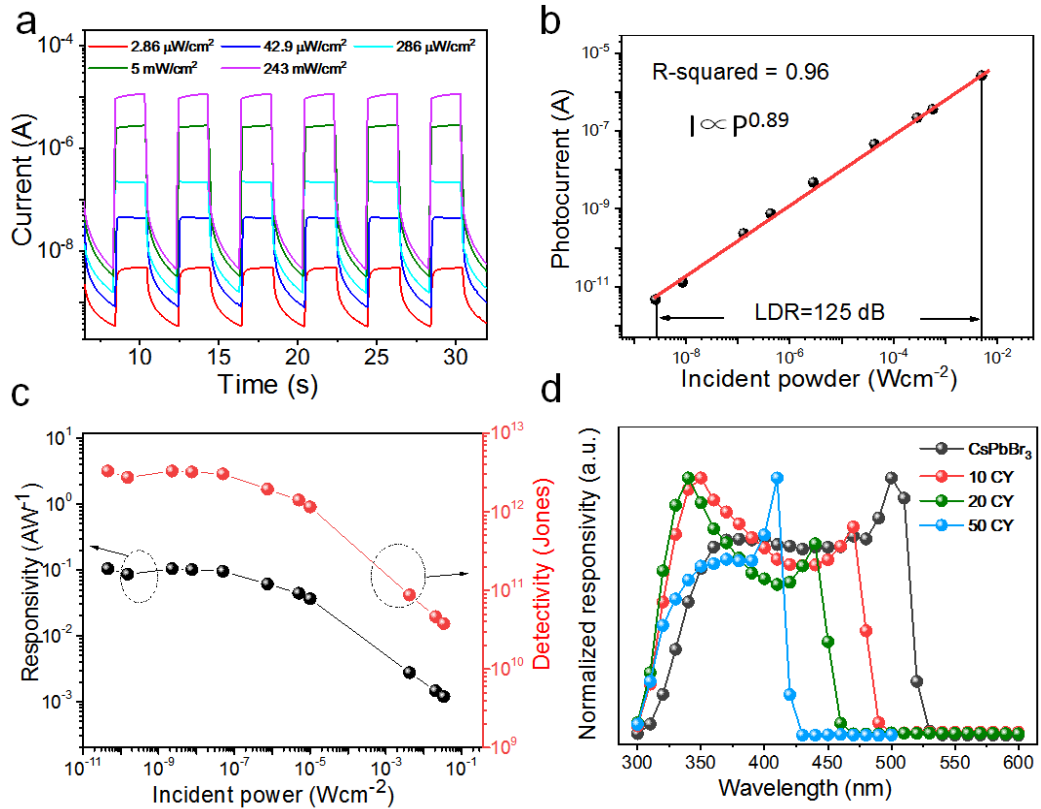

**Figure 6. UV detection performance of CsPbCl$_3$ PDs**. (a) Typical photoresponse curves under 405 nm illumination with different power density. (b) Corresponding linear dynamic range. (c) Power-dependent responsivity and specific detectivity curve, (d) photoresponse spectra by different anion-exchange process.

**Precise phase control of large-scale inorganic perovskites via vapor-phase anion-exchange strategy**

ToC figure ((Please choose one size: 55 mm broad × 50 mm high **or** 110 mm broad × 20 mm high. Please do not use any other dimensions))

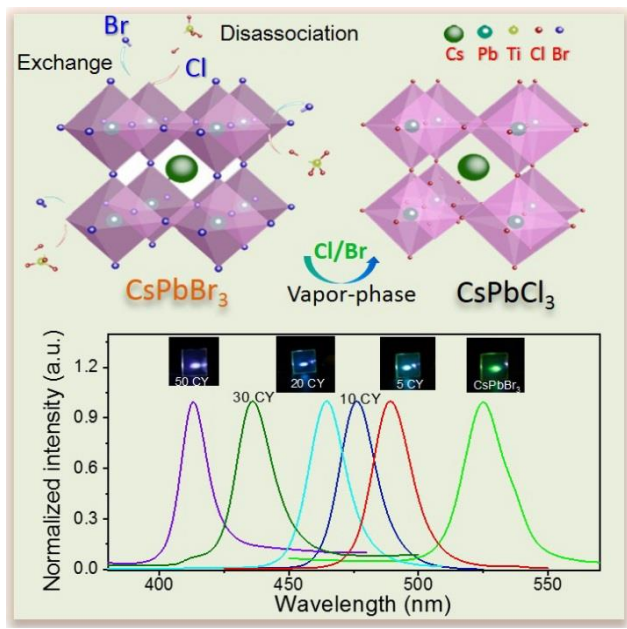

# Supporting Information

**Precise phase control of large-scale inorganic perovskites via vapor-phase anion-exchange strategy**

*Guobiao Cen[#1], Yufan Xia[#1], Chuanxi Zhao[1]\*, Yong Fu[1], Yipeng An[2], Ye Yuan[1], Tingting Shi[1]\*, Wenjie Mai[1]\**

[1]Siyuan Laboratory, Guangdong Provincial Engineering Technology Research Center of Vacuum Coating Technologies and New Energy Materials, Department of Physics, Jinan University, Guangzhou, Guangdong 510632, People's Republic of China.
[2]School of Physics & International United Henan Key Laboratory of Boron Chemistry and Advanced Energy Materials, Henan Normal University, Xinxiang, Henan 453007, China.

Contents

**Figure S1.** Solubility of perovskites precursor.

**Figure S2.** EDS characterization.

**Figure S3.** XPS spectra of Cs3d, Pb2f and Ti2p and XPS depth profile

**Figure S4.** SEM images of perovskites with different ALD cycles

**Figure S5.** The density of states (DOS) from (a) $CsPbBr_3$, (b) $CsPbBr_{1.5}Cl_{1.5}$ and (c)$CsPbCl_3$.

**Figure S6.** HSE03 projective band of (a) $CsPbBr_3$:Pb and (b) $CsPbBr_3$:Br.

**Figure S7.** HSE03 projective band of (a) $CsPbBr_{1.5}Cl_{1.5}$:Pb, (b) $CsPbBr_{1.5}Cl_{1.5}$:Br and (c) $CsPbBr_{1.5}Cl_{1.5}$:Cl.

**Figure S8.** HSE03 projective band of (a) $CsPbCl_3$:Pb and (b)$CsPbCl_3$:Cl.

**Figure S9.** GGA projective band of (a) $CsPbBr_3$:Pb and (b)$CsPbBr_3$:Br.

**Figure S10.** GGA projective band of (a) $CsPbBr_{1.5}Cl_{1.5}$:Pb, (b) $CsPbBr_{1.5}Cl_{1.5}$:Br and (c) $CsPbBr_{1.5}Cl_{1.5}$:Cl.

**Figure S11.** GGA projective band of (a) $CsPbCl_3$:Pb and (b) $CsPbCl_3$:Cl.

**Figure S12.** Absorption spectra of perovskites

**Figure S13.** General strategy for other inorganic perovskites thin films.

**Figure S14.** The detection limit and photoresponse speed of $CsPbCl_3$ UV PDs.

**Table S1.** Comparison of the calculated and reported bandgaps of $CsPbX_3$

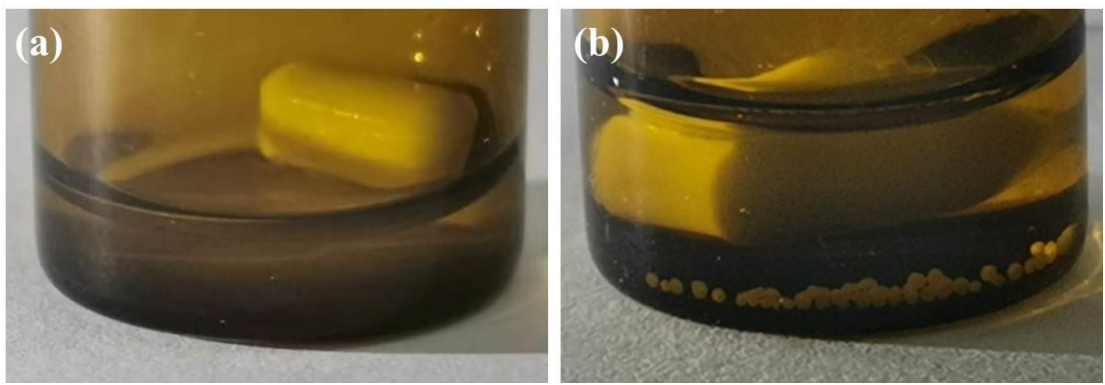

**Figure S1.** Solubility of perovskites precursors (stirred for 6 hours). (a) 0.4 M CsPbBr$_3$ precursor. (b) 0.1 M CsPbCl$_3$ precursor.

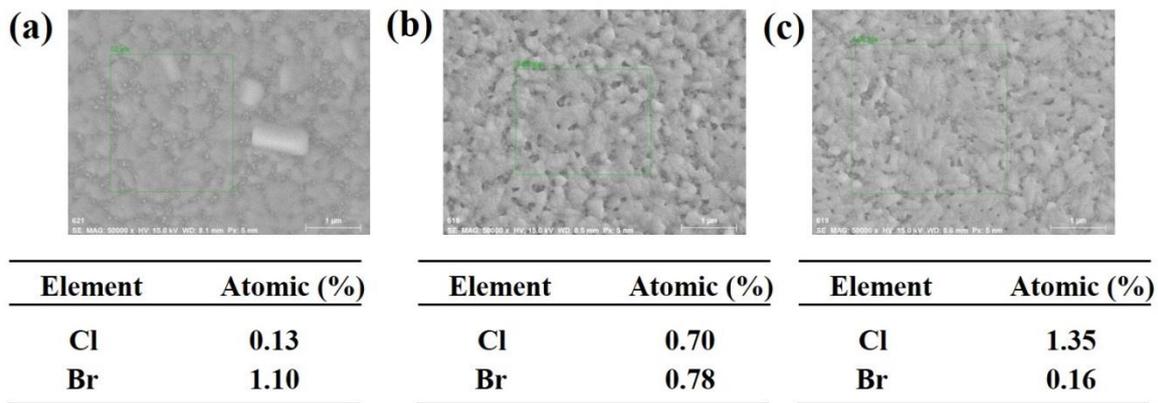

**Figure S2.** EDS of perovskites under various gas injection cycle (a) 10 cycles, (b) 20 cycles, (c) 50 cycles.

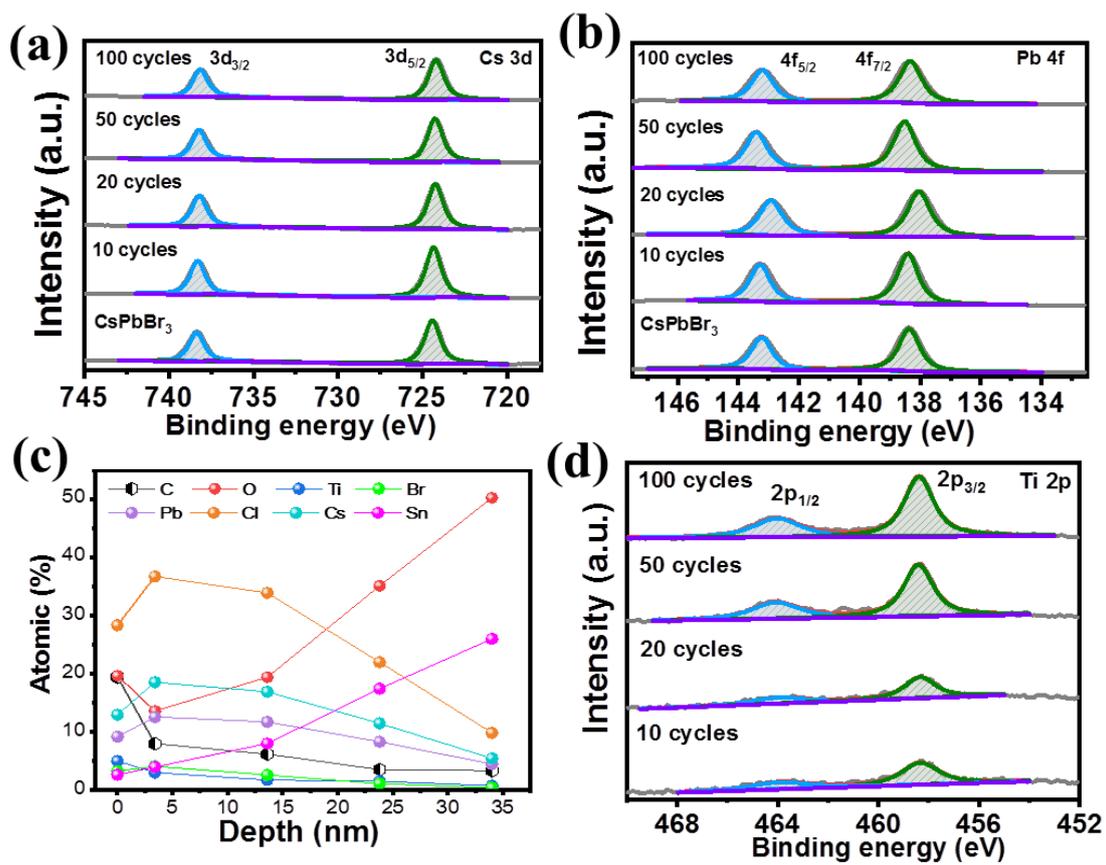

**Figure S3.** High resolution XPS spectra of (a) Cs 3d, (b) Pb 2f and (c) XPS depth profiles, (d) Ti 2p for perovskites under different ALD cycles.

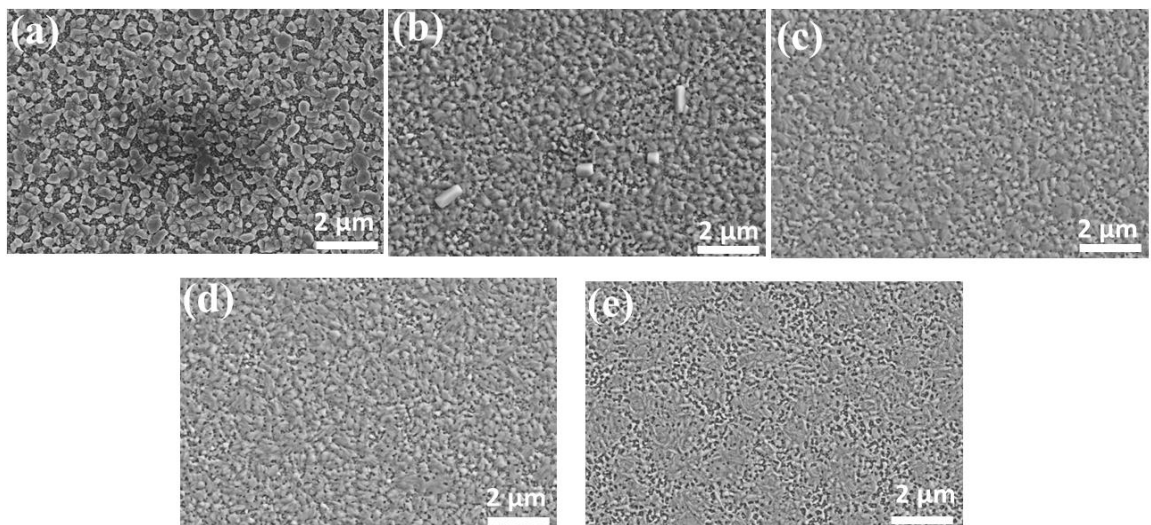

**Figure S4.** SEM images of perovskites under different gas injection cycle. (a) 0 cycle, (b) 10 cycles, (c) 20 cycles, (d) 50 cycles, (e) 100 cycles

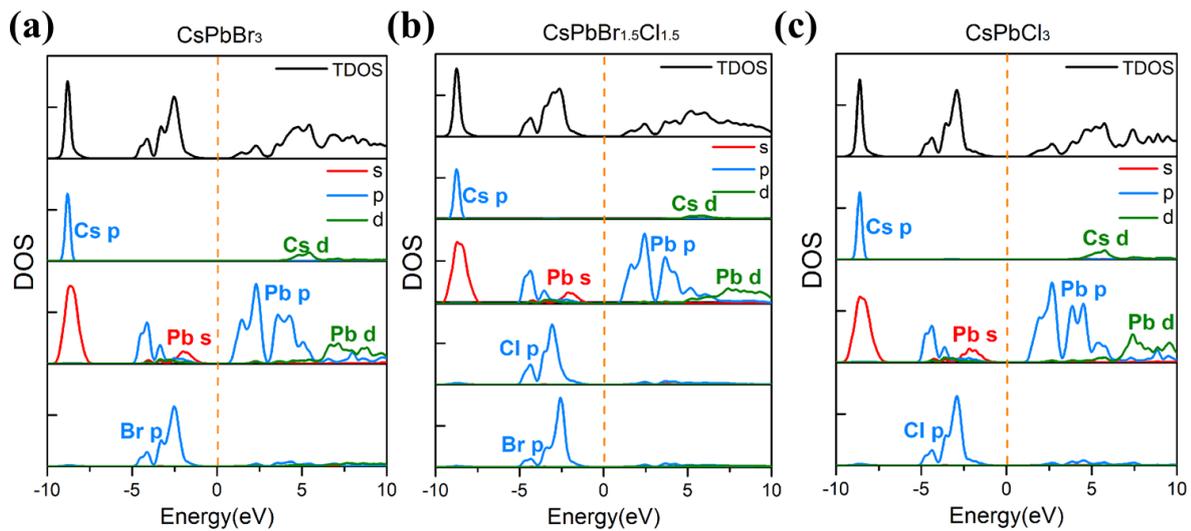

**Figure S5.** The density of states (DOS) from (a) $CsPbBr_3$, (b) $CsPbBr_{1.5}Cl_{1.5}$ and (c) $CsPbCl_3$.

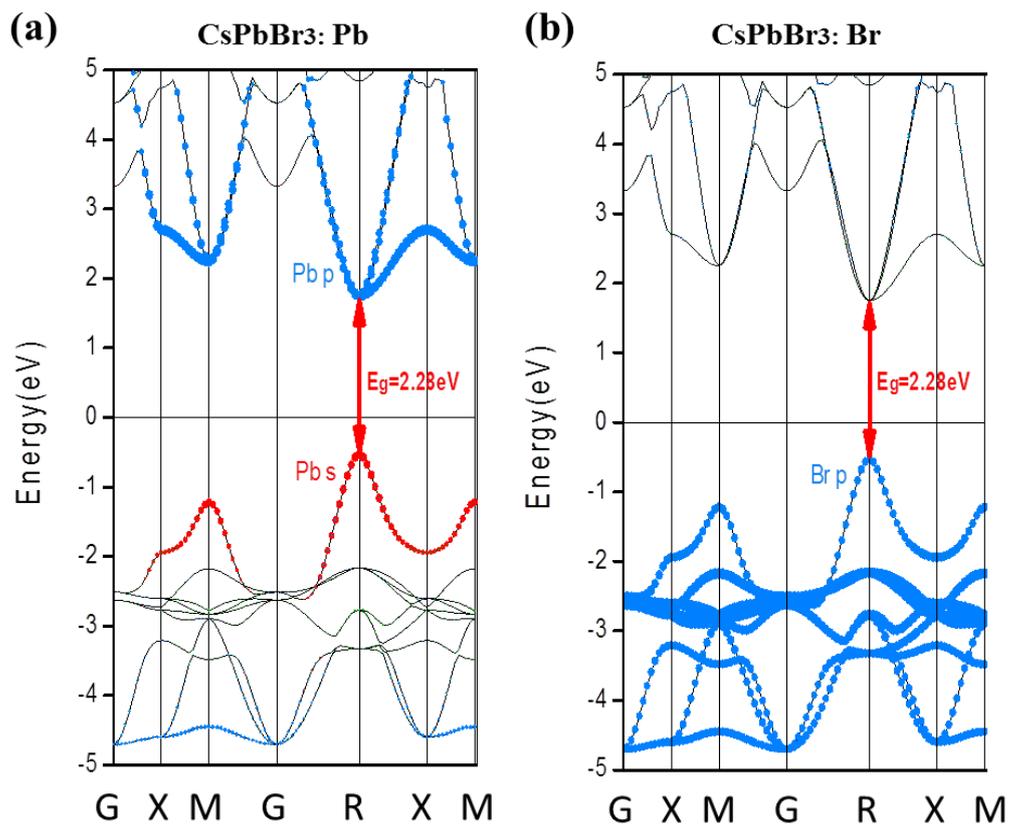

**Figure S6.** HSE03 projective band of (a) CsPbBr$_3$:Pb and (b) CsPbBr$_3$:Br

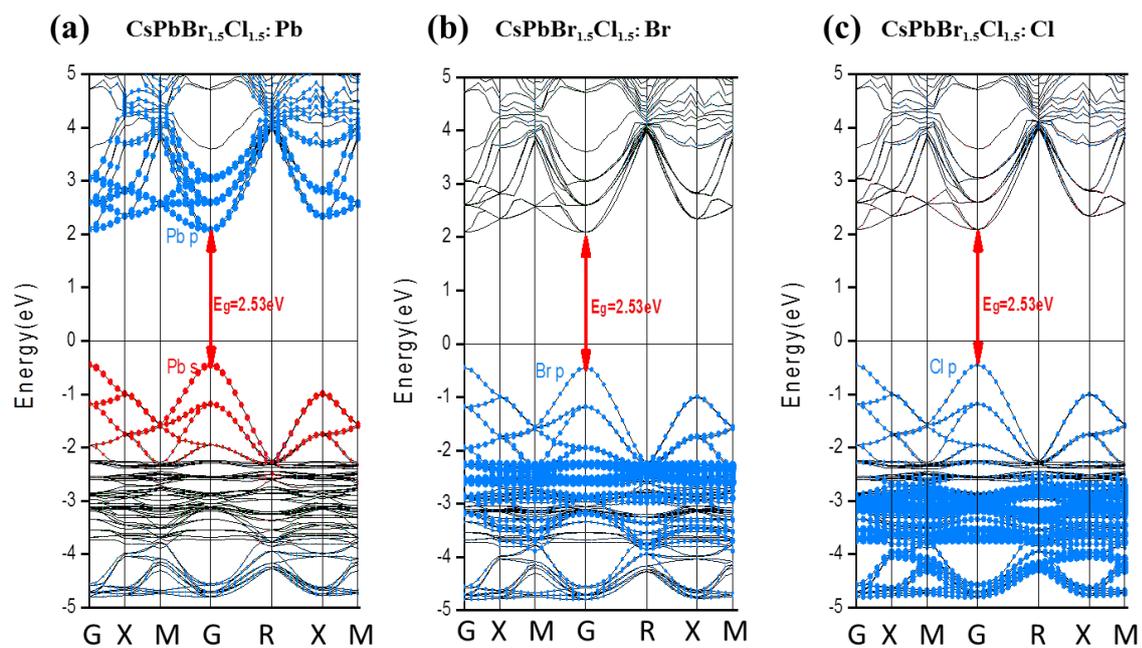

**Figure S7.** HSE03 projective band of (a) $CsPbBr_{1.5}Cl_{1.5}$:Pb, (b) $CsPbBr_{1.5}Cl_{1.5}$:Br and (c) $CsPbBr_{1.5}Cl_{1.5}$:Cl.

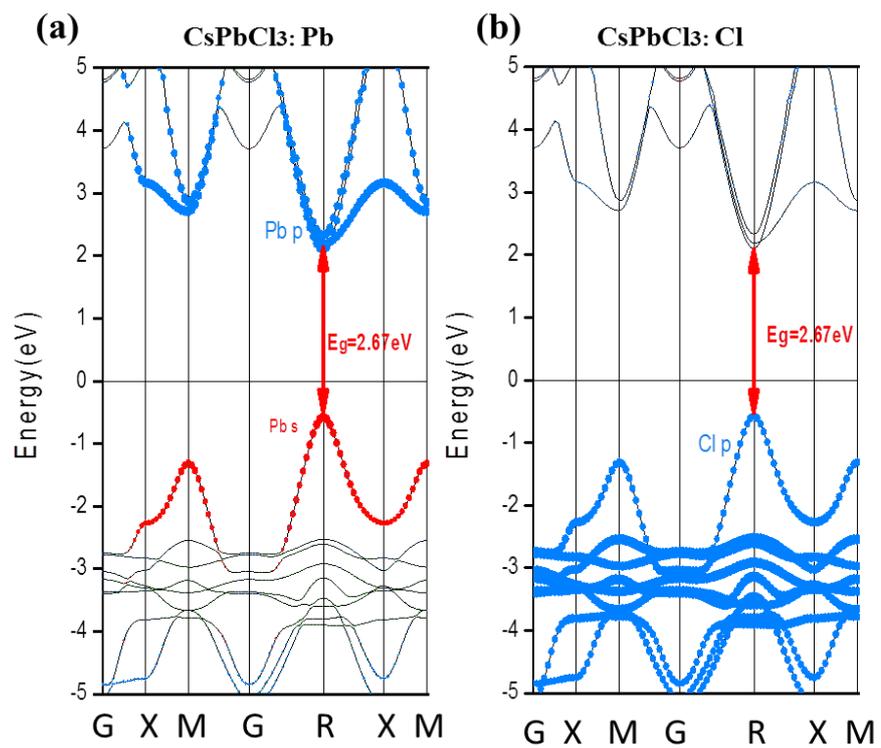

**Figure S8.** HSE03 projective band of (a) CsPbCl$_3$:Pb and (b) CsPbCl$_3$:Cl.

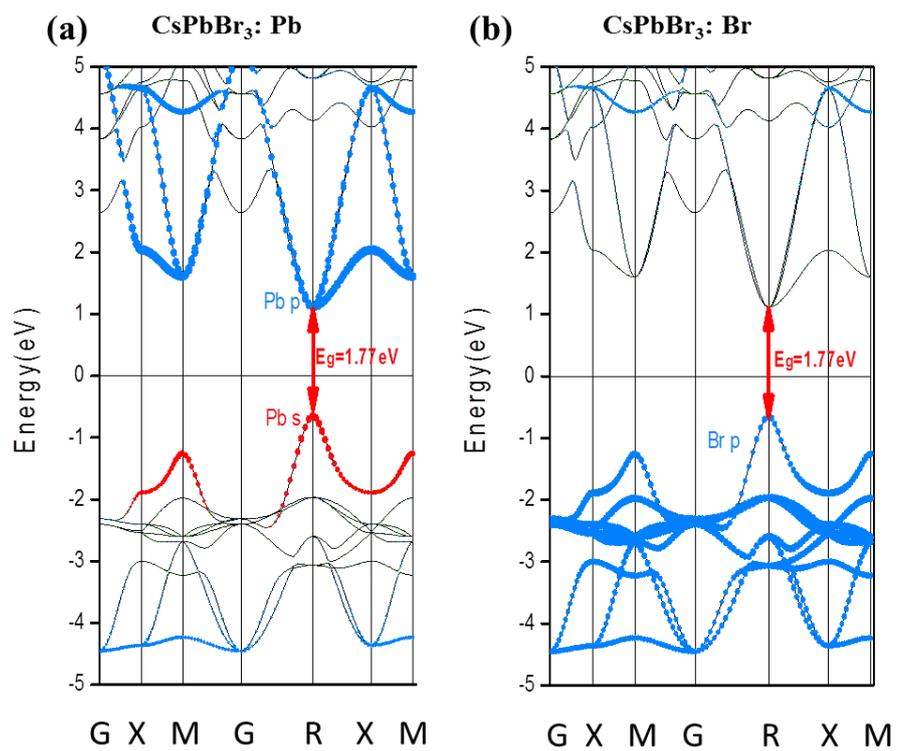

**Figure S9.** GGA projective band of (a) CsPbBr$_3$:Pb and (b) CsPbBr$_3$:Br.

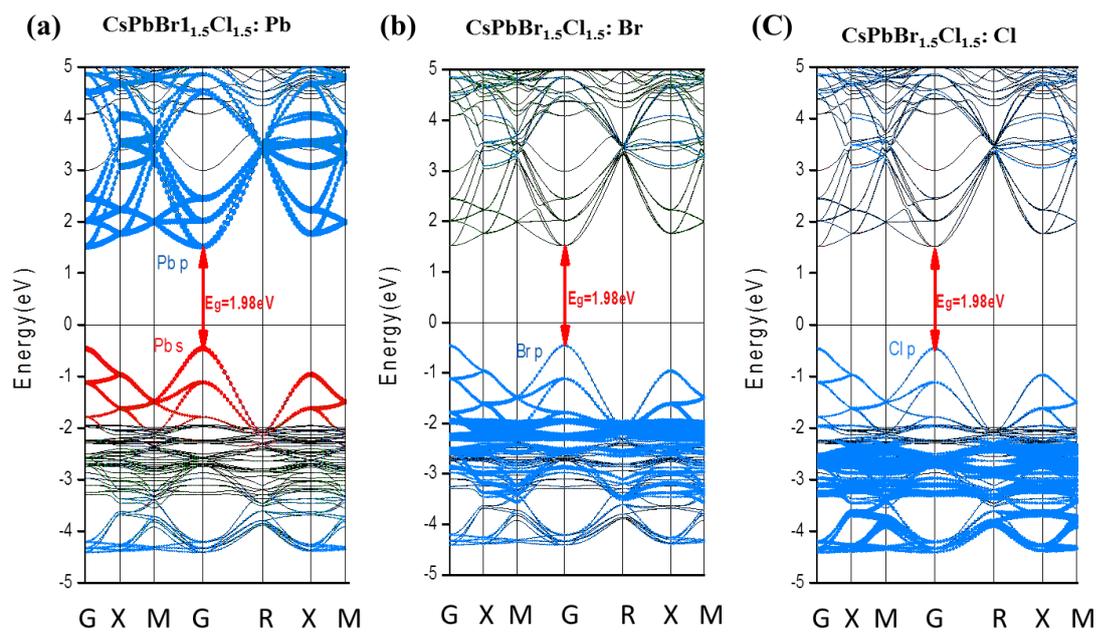

**Figure S10.** GGA projective band of (a) CsPbBr$_{1.5}$Cl$_{1.5}$:Pb, (b) CsPbBr$_{1.5}$Cl$_{1.5}$:Br and (c) CsPbBr$_{1.5}$Cl$_{1.5}$:Cl.

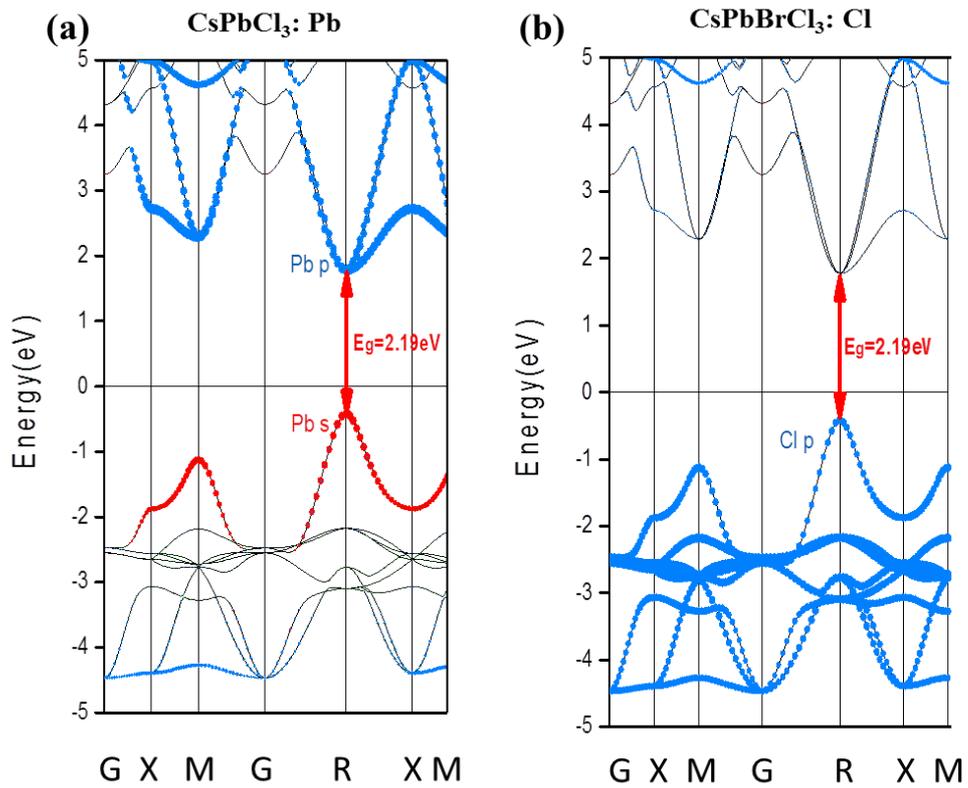

**Figure S11.** GGA projective band of (a) CsPbCl$_3$:Pb and (b) CsPbCl$_3$:Cl.

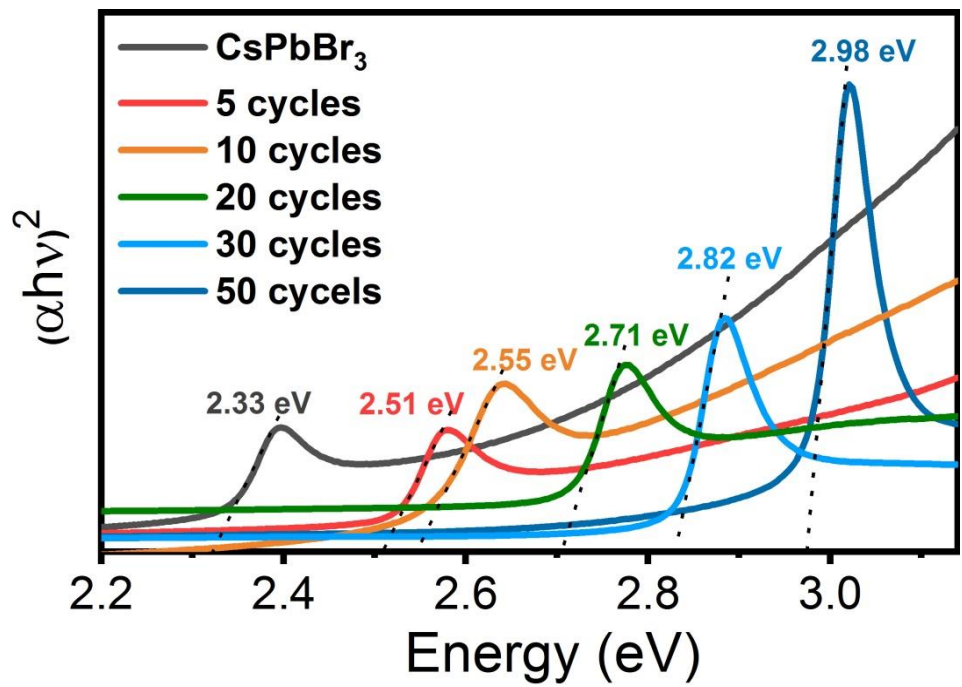

**Figure S12.** Band gaps of perovskites under different gas injection cycles.

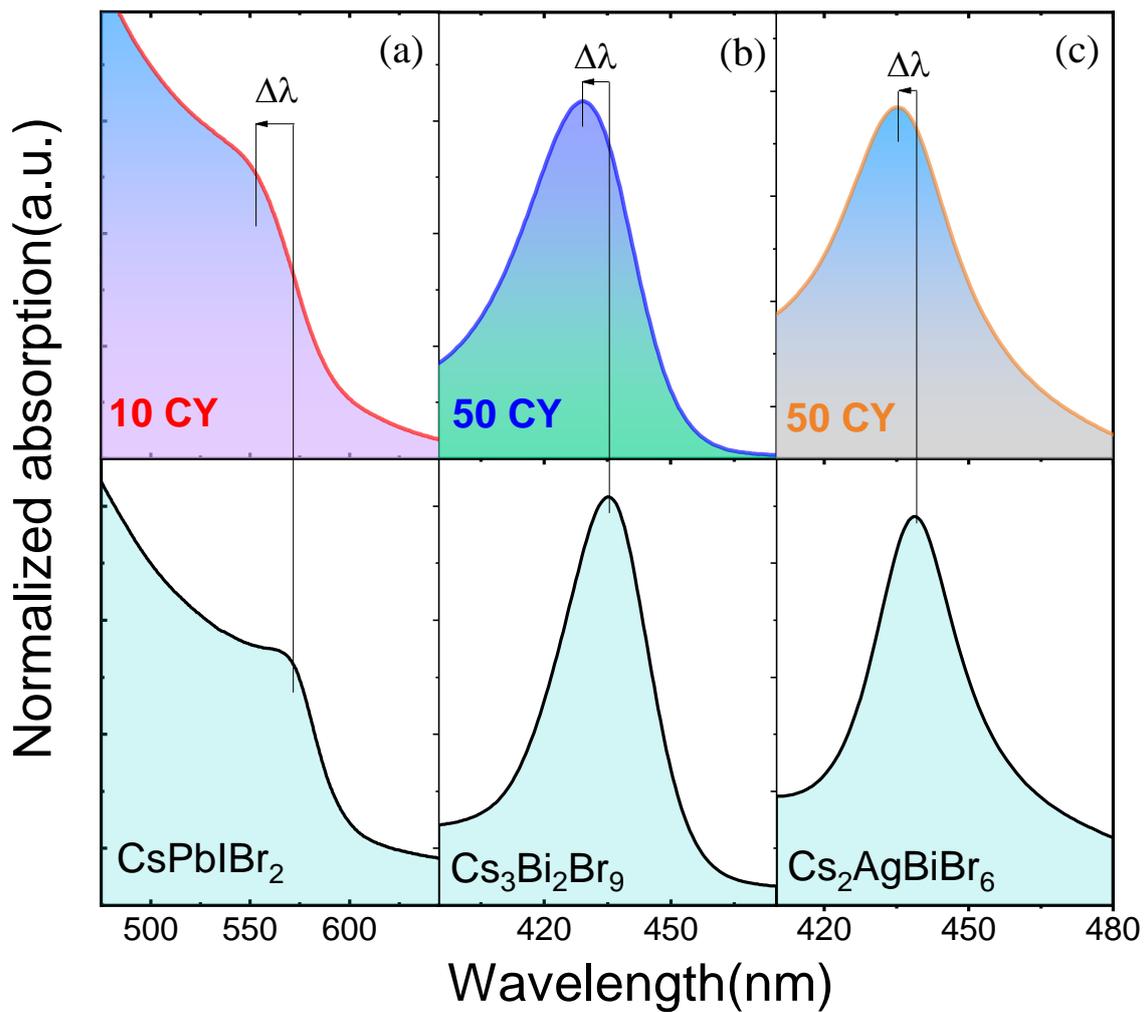

**Figure S13**. General strategy for other inorganic perovskites thin films. (a) CsPbIBr$_3$ film, (b) Cs$_3$Bi$_2$Br$_9$ film and (c) Cs$_2$AgBiBr$_6$ film.

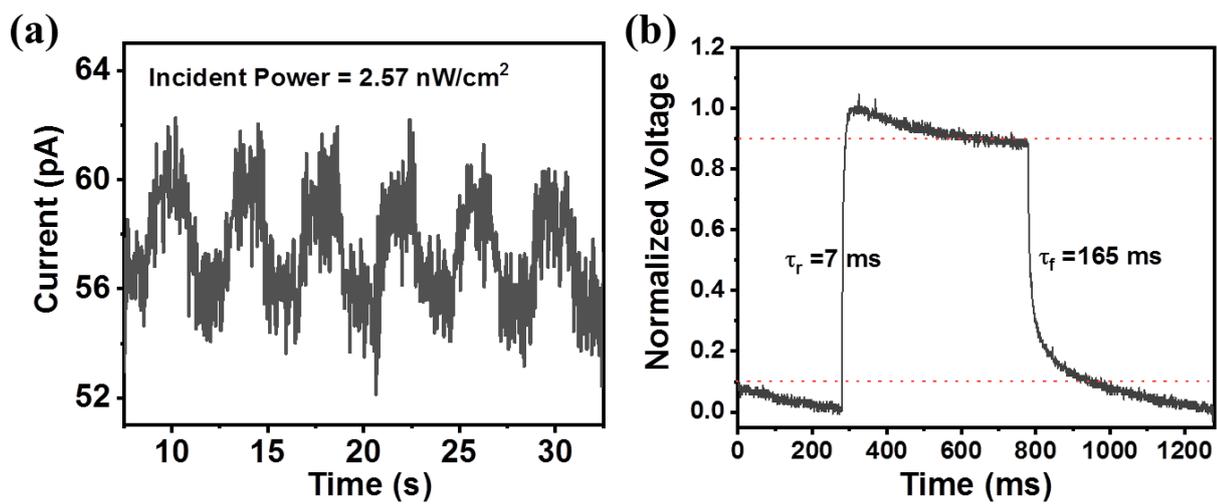

**Figure S14.** The detection limit (a) and photoresponse speed (b) of CsPbCl$_3$ UV PDs.

**Table S1. Comparison of the bandgaps of CsPbX$_3$ with previous report. Different methods but the same trends. [1]**

| Method | CsPbBr$_3$ | CsPbBr$_{1.5}$Cl$_{1.5}$ | CsPbCl$_3$ |
|---|---|---|---|
| GGA | 1.7664eV [1.764eV] | 1.9808eV | 2.1932eV [2.172eV] |
| GGA+SOC | 0.4380eV | 0.6515eV | 0.8578eV |
| PBE0 | 3.0919eV | 3.2941eV | 3.4671eV |
| PBE0+SOC | 1.2742eV | 1.4814eV | 1.7552eV |
| HSE03 | 2.2809eV | 2.5339eV | 2.6710eV |
| HSE06 | 2.4606eV | 2.7227eV | 2.8417eV |

[1] Ahmad M, Rehman G, Ali L, Shafiq M, Iqbal R, Ahmad R, *et al.* Structural, electronic and optical properties of CsPbX$_3$ (X=Cl, Br, I) for energy storage and hybrid solar cell applications. *Journal of Alloys and Compounds* 2017, **705:** 828-839.